\documentclass[12pt]{article}
\pdfoutput=1
\usepackage{amsmath,amssymb}
\usepackage{graphicx}
\usepackage{color}
\usepackage[bookmarksnumbered=true,colorlinks=true,linkcolor=black,citecolor=black]{hyperref}
\newlength{\extraspace}
\newlength{\extraspaces}
\def\bsklength{.8mm} 

\newcommand{\beq}{\begin{equation}}
\newcommand{\eeq}{\end{equation}}
\newcommand{\bseq}{\addtocounter{subeqno}{1}\begin{subequations}}
\newcommand{\eseq}{\end{subequations}}

\font\mathscript=eusm10 at 12pt
\font\mathscripts=eusm7
\font\mathscriptss=eusm5
\newfam\mathscri
\textfont\mathscri=\mathscript
\scriptfont\mathscri=\mathscripts
\scriptscriptfont\mathscri=\mathscriptss
\def\mathscr#1{{\fam\mathscri\relax#1}}

\font\mathfrakt=eufm10 at 12pt
\font\mathfrakts=eufm7
\font\mathfraktss=eufm5
\newfam\mathfraki
\textfont\mathfraki=\mathfrakt
\scriptfont\mathfraki=\mathfrakts
\scriptscriptfont\mathfraki=\mathfraktss
\def\mathfrak#1{{\fam\mathfraki\relax#1}}

\def\CH{{\cal H}}

\def\CO{{\cal O}}

\renewcommand{\tilde}{\widetilde}

\renewcommand{\bar}{\overline}

\def\half{{\textstyle\frac{1}{2}}}

\newcommand{\lsim}{\lesssim}

\begin{document}
\setcounter{page}{0}
\addtolength{\baselineskip}{\bsklength}
\thispagestyle{empty}
\renewcommand{\thefootnote}{\fnsymbol{footnote}}        

\begin{flushright}
arXiv:yymm.nnnn [hep-ph]\\
\end{flushright}
\vspace{.4cm}

\begin{center}
{\Large
{\bf{Flavor Violating Lepton Family U(1)$_\lambda$}}}\\[1.2cm]
{\rm HoSeong La\footnote{hsla.avt@gmail.com}
}
\\[3mm]
{\it Department of Physics and Astronomy,\\[1mm]
Vanderbilt University,\\[1mm]              
Nashville, TN 37235, USA} \\[1.5cm]

\vfill
{\parbox{15cm}{
\addtolength{\baselineskip}{\bsklength}
\noindent
The Standard Model is extended minimally with
a new flavor-violating family symmetry ${\rm U(1)}_\lambda$,
which acts only on leptons including the right-handed neutrinos.
The model is anomaly free with family-dependent ${\rm U(1)}_\lambda$ charges,
and consistent with the observed neutrino mixing angles. 
It predicts charged lepton flavor-violating processes mediated by 
a new gauge boson.
Under certain conditions, the smallness of $\theta_{13}$ of neutrino
mixing can be justified in terms of the muon-to-tau mass ratio, at the same 
time explaining the electron-to-tau large mass hierarchy.

\bigskip
}
}


\end{center}
\noindent
\vfill


\newpage
\setcounter{page}{1}
\setcounter{section}{0}
\setcounter{equation}{0}
\setcounter{footnote}{0}
\renewcommand{\thefootnote}{\arabic{footnote}}  
\newcounter{subeqno}
\setcounter{subeqno}{0}
\setlength{\parskip}{2mm}
\addtolength{\baselineskip}{\bsklength}

\pagenumbering{arabic}


\noindent
\underline{Introduction}

The observations of the neutrino flavor violation indicate the existence of 
new physics beyond the Standard Model (SM)\cite{pdg}. 
This is usually interpreted in terms of the neutrino oscillations 
based on the mismatch between the flavor and mass 
eigenstates\cite{Pontecorvo:1957qd}\cite{no_ref}. However, if neutrino masses
are originated from some kind of Higgs mechanism, the theory must include
tree-level neutrino flavor violating Yukawa interactions. Furthermore, 
this neutrino flavor violation would lead to charged lepton flavor 
violations (cLFV) at one-loop with a $W$-boson exchange. 
Although this induced cLFV is negligibly small, but it inevitably raises 
a question that if there is a Beyond-SM(BSM) with more significant and
explicit lepton flavor violating interactions including charged leptons 
(see \cite{Hoecker:2012np}\cite{Deppisch:2012vj} for recent reviews).

In this Letter, we will explore the possibility how the right-handed
neutrinos can play the role of going beyond the SM.
One way is to introduce some gauge charge explicitly for them\cite{La:2012ky}.
So we will consider an extra ${\rm U(1)}_\lambda$ gauge interaction, 
which acts solely on leptons including the right-handed neutrinos. 
${\rm U(1)}_\lambda$ has different charges for different
eigenstates similarly as in the Froggatt-Nielsen case\cite{Froggatt:1978nt}, 
which leads to a flavor violating interaction in the 
flavor basis\cite{La:2012ky}. 
Then we can investigate the physics of cLFV, in particular, generated by 
a gauge interaction beyond Yukawa interactions. 
Note that our proposal is different from
gauging the lepton number as a gauge symmetry, which was first suggested 
in \cite{Foot:1989ts}. Recent references include 
\cite{FileviezPerez:2010gw}\cite{Chao:2010mp}\cite{Schwaller:2013hqa},
but all in the flavor conserving context.

\noindent
\underline{The Model}

The model we consider includes the right-handed neutrinos and 
the gauge group is 
\beq
\label{e:1}
G ={\rm SU}(3)_C\times {\rm SU}(2)_L\times{\rm U}(1)_Y
\times {\rm U}(1)_\lambda.
\eeq
${\rm U}(1)_\lambda$ acts only on leptons including $\nu_{\lambda R}$
with charges $q_\lambda$ depending on the flavor generation. This also allows
$\nu_{\lambda R}$ to be genuine right-handed partners of left-handed neutrinos.
Although we expect that ${\rm U}(1)_\lambda$ is broken at a scale near or
above the Electroweak (EW) scale, the coupling constant can be very small 
so that its gauge boson can couple to leptons very weakly.
We assume that ${\rm U(1)}_\lambda$ is left-right (LR) symmetric to minimize 
new anomalies, but this still is not enough to cancel all anomalies.
For example, $({\rm SU(2)})^2{\rm U(1)}_\lambda$ anomaly does not cancel 
unless the sum of ${\rm U(1)}_\lambda$ charges for all generations cancel. 
In fact, the condition 
\beq
\label{e:1a}
\sum q_\lambda =0,
\eeq
where $q_\lambda$ are diagonalized ${\rm U(1)}_\lambda$ charges for 
${\rm U(1)}_\lambda$ eigenstates,
is the only extra we need to cancel all anomalies as shown below:
\bseq
\begin{align}
({\rm SU(2)})^2{\rm U(1)}_\lambda &\propto \sum q_\lambda, \\
({\rm U(1)}_\lambda)^3 &\propto \sum_L (q_\lambda)^3 -\sum_R (q_\lambda)^3=0, \\
({\rm U(1)}_\lambda)^2{\rm U(1)}_Y &\propto 
\left(2(-\half)-(-1)\right)\sum (q_\lambda)^2=0,\\
({\rm U(1)}_Y)^2{\rm U(1)}_\lambda &\propto 
\left(2(-\half)^2-(-1)^2\right)\sum q_\lambda,\\
(\mbox{Gravity})^2{\rm U(1)}_\lambda 
&\propto \sum_L q_\lambda -\sum_R q_\lambda=0,
\end{align}
\eseq
where all anomalies are summed over all generations since they do not 
necessarily cancel separately in each generation.
This structure clearly indicates the ${\rm U(1)}_\lambda$ is not a gauged
lepton number symmetry.

Extra U(1) is very common in 
extended SM\cite{Langacker:2008yv}\cite{Langacker:2009im}. 
(Also see \cite{Holdom:1985ag}.)
Other models that accommodate family-dependent U(1) family symmetry
include \cite{Ibanez:1994ig}, but flavor violation is from the Yukawa sector,
not gauge sector. Also these models are not left-right symmetric so that 
the anomaly cancellation is more complicated, some even possible only by 
inheriting from the Green-Schwarz mechanism in string theory. 
Their extra U(1) acts on both quarks and leptons.
So none of these is similar to our model. The closest is 
\cite{Demir:2005ti}, however, which did not consider the possibility of
LFV. \cite{Langacker:2000ju} and \cite{Chen:2006hn} consider 
the flavor changing neutral current (FCNC) due to $Z'$ by recognizing that
the $Z'$ couplings to fermions can be non-diagonal, 
but the extra U(1) acts on quarks as well. So, it is different from our model.

\noindent
\underline{Generic Structure and Notations}

Some comments on the notations are in order.
We will use, generically, $\psi_\ell$ to denote the (Weak) flavor eigenstates, 
$\psi_j$ mass eigenstates, and $\psi_\lambda$ ${\rm U(1)}_\lambda$ eigenstates.
More specifically, $\ell$ ($\nu_\ell$) stands for flavor eigenstates 
for charged leptons (neutrinos, respectively), and
$\lambda$ ($\nu_\lambda$) stands for ${\rm U(1)}_\lambda$ eigenstates
for charged leptons (neutrinos, respectively).
For charged leptons, eventually we will identify the mass and flavor eigenstates 
such that $\ell$ stands for physical states $e, \mu, \tau$.

In terms of the (generic) flavor eigenstates $\psi_\ell$, 
the ${\rm U(1)}_\lambda$ coupling to leptons are of the form
\beq
\label{e:2}
\CH_{\rm int}^{(\lambda)}
\equiv\bar{\psi_{\ell'}}\gamma^\mu q_{\ell'\ell} Z^{(\lambda)}_\mu\psi_{\ell},
\eeq
where $Z^{(\lambda)}_\mu$ is the ${\rm U(1)}_\lambda$ gauge boson.
If the hermitian matrix
\beq
\mathbf{q}\equiv (q_{\ell'\ell})
\eeq
is not diagonal, we have LFV.
The ${\rm U(1)}_\lambda$ gauge invariance can be shown in terms of the 
${\rm U(1)}_\lambda$ eigenstates $\psi_\lambda$, 
for which the ${\rm U(1)}_\lambda$ 
gauge coupling constant matrix $\mathbf{q}_d$ becomes diagonal.
Since $\mathbf{q}$ is a matrix in the flavor space,
the corresponding matrix $\mathbf{q}_d$ in the ${\rm U(1)}_\lambda$ eigenstate
basis can be generated by a unitary transformation
\beq
\label{e:3}
{q}_{d\lambda'\lambda}
=(V^\dagger)_{\lambda'\ell'}q_{\ell'\ell} V_{\ell \lambda},
\eeq
where $V$ is given by
\beq
\label{e:4}
\psi_\ell= V_{\ell \lambda}\psi_\lambda,\quad V^\dagger V=1
\eeq
and parametrized\footnote{We will ignore CP-violations, 
for simplicity, and all transformation matrices are parametrized similarly.}
\beq
\label{e:r14}
V({\tilde{\theta}}_{12},{\tilde{\theta}}_{23},{\tilde{\theta}}_{13})=
\left(
\begin{array}{ccc}
1 & 0 &0\\ 
0 &\tilde{c}_{23}  &\tilde{s}_{23}\\
0 &-\tilde{s}_{23} &\tilde{c}_{23}
\end{array}
\right)
\left(
\begin{array}{ccc}
\tilde{c}_{13} & 0 &\tilde{s}_{13}\\ 
0 &1  &0\\
-\tilde{s}_{13} &0 &\tilde{c}_{13}
\end{array}
\right)
\left(
\begin{array}{ccc}
\tilde{c}_{12} & \tilde{s}_{12} &0\\ 
-\tilde{s}_{12} &\tilde{c}_{12}  &0\\
0 &0 &\ 1
\end{array}
\right),
\eeq
where $\tilde{c}\equiv \cos\tilde{\theta}$ and 
$\tilde{s}\equiv \sin\tilde{\theta}$.
Note that this unitary transformation does not affect any of the SM 
gauge couplings since they are all family independent, i.e. proportional 
to identity in the flavor basis.
The anomaly cancellation requires that the coupling constant matrix be 
traceless. So, in the ${\rm U(1)}_\lambda$ basis,
\beq
\label{e:s0}
\mathbf{q}_d
={\rm diag}(q_1, q_2, -q_1-q_2).
\eeq
There are three distinctively different cases: $q_1 =0$, 
$q_2=q_1$, or $q_2\neq q_1\neq 0$.

The charged lepton masses in the ${\rm U(1)}_\lambda$ basis are given by
\beq
\label{e:5}
\mathbf{M}=(M_{\lambda'\lambda})=V^{-1} \mathbf{M}_d V,
\eeq
where $\mathbf{M}_d$ is, with the identification of 
the mass and flavor eigenstates for charged leptons, the diagonalized
mass matrix given in terms of the physical masses as
\beq
\label{e:6}
\mathbf{M}_d={\rm diag}(M_\ell)={\rm diag}(M_e, M_\mu, M_\tau).
\eeq
The Dirac neutrino mass matrix in the ${\rm U(1)}_\lambda$ basis, 
$\mathbf{m}$, is not necessarily diagonalized in the same way as $\mathbf{M}$,
but as
\beq
\label{e:8}
\mathbf{m}_d={\rm diag}(m_{Di})= U_D^{-1} \mathbf{m} U_D.
\eeq
The right-handed Majorana mass matrix $\tilde{\mathbf{m}}_R$ 
in the ${\rm U(1)}_\lambda$ basis is diagonalized 
as
\beq
\label{e:7}
\tilde{\mathbf{m}}_{Rd}
={\rm diag}(\tilde{m}_{Ri})=U_R^{-1}\tilde{\mathbf{m}}_R U_R.
\eeq
Since ${\rm U(1)}_\lambda$ gauge couplings to $\nu_L$ and $\nu_R$
are supposed to be the same in any basis,
\beq
\label{e:7a}
U_D=U_R.
\eeq
To satisfy this, neutrino mass matrices must commute with each other as
\beq
\label{e:9}
[\mathbf{m}, \tilde{\mathbf{m}}_R]=0.
\eeq
If $\tilde{\mathbf{m}}_R\neq 0$, the physical neutrino masses are 
necessarily Majorana for both chiralities and 
given in eqs.(\ref{e:s16c})(\ref{e:s16d}).
Since we identify the flavor and mass eigenstates for charged leptons
such that $U_\ell^\dagger =1$, from eqs.(\ref{e:4})(\ref{e:8} or \ref{e:7}),
the neutrino mixing matrix $U_\nu$ can be identified as
\beq
\label{e:10}
U_{\rm PMNS}=U_\nu
=V U_D
= V U_R.
\eeq
Note that the r.h.s has six angles, while the l.h.s has only three.
For the best-fit values, we take the latest PDG numbers\cite{pdg2013}:
$\sin^2(2\theta_{12})=0.857{+0.023\atop -0.025}$, 
$\sin^2(2\theta_{23})>0.95$,
and $\sin^2(2\theta_{13})=0.095\pm 0.01$. Then the best-fit
values for the mixing matrix we can use are
\beq
\begin{aligned}
\label{e:s4bf}
\sin^2\theta_{12} &\simeq 0.311 \pm 0.016,\\ 
\sin^2\theta_{23}&\simeq (0.39\ {\sim}\ 0.61), \\
\sin^2\theta_{13}&\simeq 0.024 \pm 0.003.
\end{aligned}
\eeq

Without ${\rm U(1)}_\lambda$ charged doublet $\Phi_\lambda$, 
we have twelve free parameters ($q_1$, $q_2$, momentum cut-off $\Lambda$, 
and nine diagonal Yukawa coupling constants) plus the number of 
vacuum expectation values $v_\lambda$ of the symmetry breaking scalar fields, 
but only eight constraints (three charged lepton masses, 
three neutrino mixing angles, and two neutrino mass relations).
We can assume $\Lambda\sim v_\lambda$, but this still leaves at least four
unconstrained parameters. The rest of details depend on the choices of 
these free parameters. With $\Phi_\lambda$, the number increases because of
non-diagonal Yukawa coupling constants.

\noindent
\underline{Dirac Masses for Charged Leptons}

The Yukawa couplings for charged leptons are given by
\beq
\label{e:s5a}
Y_{\lambda'\lambda}\bar{L_{\lambda' L}} H \ell_{\lambda R}+{\rm h.c.},
\eeq
where $H$ is the SM isospin doublet Higgs and $L$'s are the lepton doublets. 
Upon the EW symmetry breaking, this leads to Dirac mass terms for charged leptons as
\beq
\label{e:11}
\bar{\lambda'}M_{\lambda'\lambda}\lambda.
\eeq
Since the SM Higgs does not carry ${\rm U(1)}_\lambda$ charge, 
the ${\rm U(1)}_\lambda$ gauge invariance requires that the tree-level 
SM Higgs Yukawa couplings should be diagonal in the ${\rm U(1)}_\lambda$ basis
or some charges are degenerate\footnote{Subsequently, we will assume charges are non-degenerate since, otherwise, it does not lead to cLFV.}. 
Without any other off-diagonal contributions, this would imply that charged 
leptons have the same mass and ${\rm U(1)}_\lambda$ charge eigenstates so that 
${\rm U(1)}_\lambda$ would not induce cLFV. 

\begin{figure}[t] 
\begin{minipage}[b]{0.48\linewidth}
\centering
\includegraphics[width=2in]{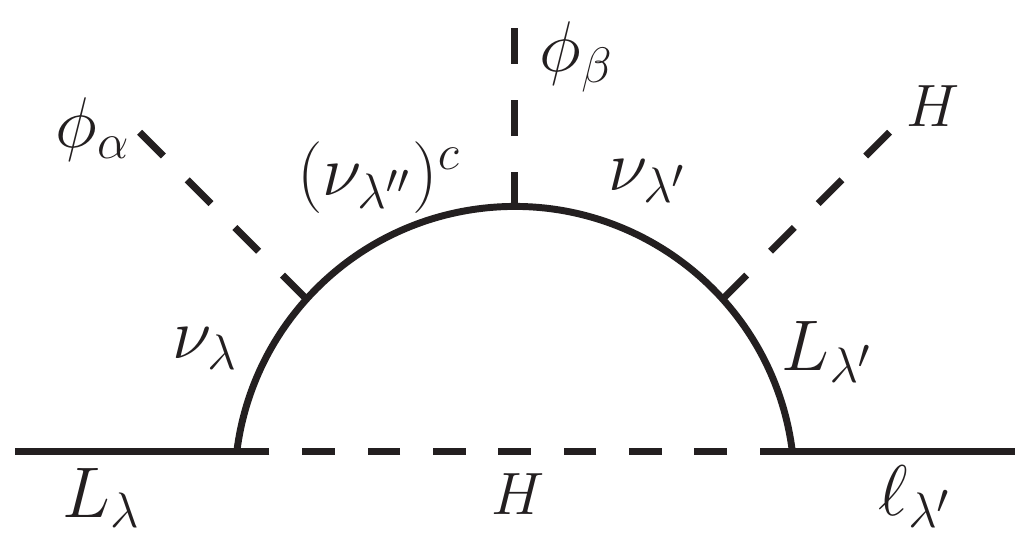} 
   \parbox{0.9\textwidth}{\vskip-6pt
   \caption{One-loop contributions to charged lepton masses. Each vertex must
   conserve ${\rm U(1)}_\lambda$ charges.}
   \label{fig:1a}}
\end{minipage}
\begin{minipage}[b]{0.48\linewidth}
\centering
   \includegraphics[width=2in]{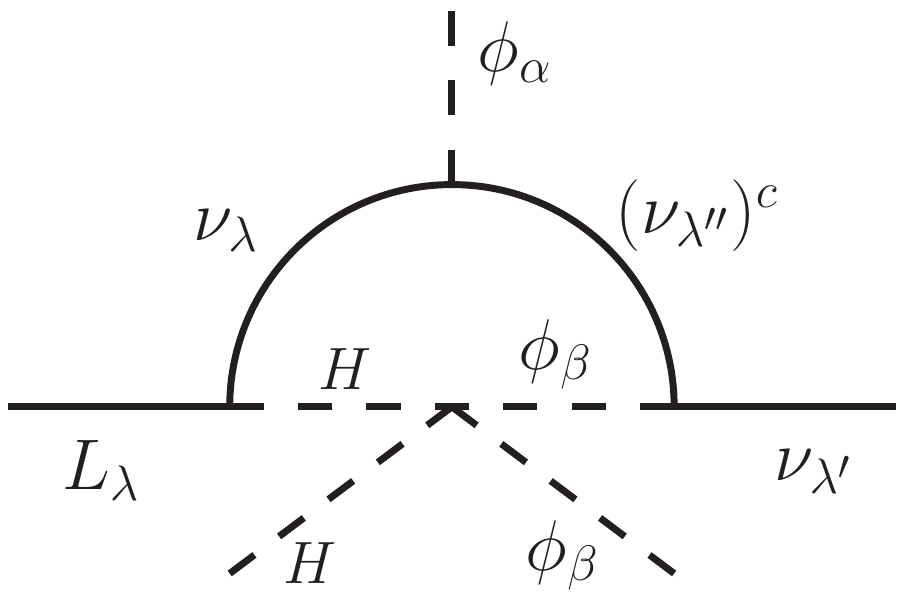} 
   \parbox{0.9\textwidth}{\vskip-6pt
   \caption{One-loop contributions to Dirac neutrino masses.}
   \label{fig:2}}
\end{minipage}
\end{figure}

However, this is not true due to the Majorana Yukawa couplings of $\nu_R$
with the ${\rm U(1)}_\lambda$ breaking scalars (see eq.(\ref{e:s6})), 
which will generate non-diagonal Dirac mass terms for leptons at higher 
orders.\footnote{Off-diagonal mass terms can also be generated if 
${\rm U(1)}_\lambda$ charged isospin doublet scalar $\Phi_\lambda$ is 
introduced.}
For example, the leading order of non-diagonal Dirac masses for charged leptons 
can be generated at one-loop, shown in Fig.\ref{fig:1a}, as
\beq
\label{e:r16x}
\Delta M_{\lambda'\lambda}^{(1)}
\sim
\sum_{\alpha,\beta} {v_\alpha v_\beta v_{\rm EW}\over \Lambda^2}\,
Y_{\lambda'\lambda'}\, y_{\lambda\lambda}\, y_{\lambda'\lambda'}\,
\tilde{y}_{\lambda\lambda''}\, \tilde{y}_{\lambda'\lambda''}
+(\lambda\leftrightarrow\lambda'),
\eeq
where Yukawa couplings at tree-levels are determined by
\beq
\label{e:r16}
Y_{\lambda\lambda}{v_{\rm EW}\over\sqrt{2}} =M_{\lambda\lambda}^{(0)},\qquad
y_{\lambda'\lambda'}{v_{\rm EW}\over\sqrt{2}} =m_{\lambda'\lambda'}^{(0)},\qquad
\tilde{y}_{\lambda\lambda''}v_\alpha =\tilde{m}^{(0)}_{R\lambda\lambda''}.
\eeq

\noindent
\underline{Dirac Masses for Neutrinos}

The SM Higgs also has Yukawa couplings for neutrinos as
\beq
\label{e:s5b}
y_{\lambda\lambda}\bar{L_{\lambda L}} i\sigma_2 H^* \nu_{\lambda R},
\eeq
and the most general (non-diagonal) Dirac masses for neutrinos are
\beq
\label{e:h3}
\bar{\nu_{\lambda'}}m_{\lambda'\lambda}\nu_\lambda,
\eeq
with including non-diagonal Dirac neutrino masses generated 
at one-loop, shown in Fig.\ref{fig:2}, as
\beq
\label{e:h2}
\Delta m_{\lambda'\lambda}^{(1)}
\sim
\sum_{\alpha,\beta} {v_\alpha v_\beta v_{\rm EW}\over \Lambda^2}\,
y_{\lambda\lambda}\,
\tilde{y}_{\lambda\lambda''}\, \tilde{y}_{\lambda'\lambda''}\, 
\lambda_{H\phi},
\eeq
which is larger than those from Fig.\ref{fig:1a} with $\ell_{\lambda'}$ 
replaced by $\nu_{\lambda'}$ if $\lambda_{H\phi}$ is not too small.
This makes it possible that the off-diagonal masses of neutrino Dirac
mass matrix are relatively larger even if those
of charged lepton mass matrix are much smaller than the diagonal values.

\noindent
\underline{Majorana Masses for $\nu_R$}

\begin{figure}[t] 
\begin{minipage}[t]{0.49\linewidth}
\centering
   \includegraphics[width=2in]{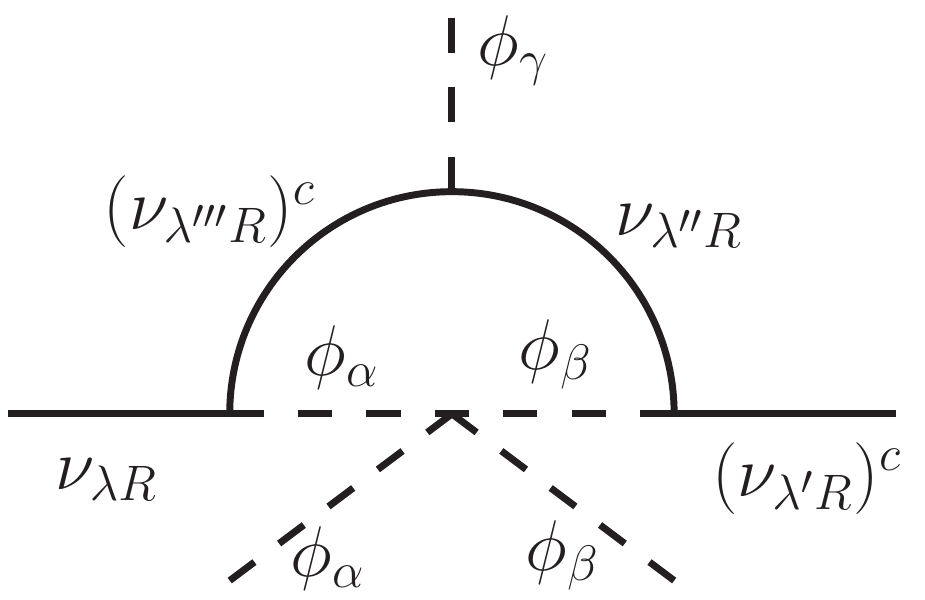} 
   \parbox{0.95\textwidth}{\vskip-6pt
   \caption{One-loop contributions to Majorana neutrino masses}
   \label{fig:3a}}
\end{minipage}
\begin{minipage}[t]{0.49\linewidth}
\centering
   \includegraphics[width=2in]{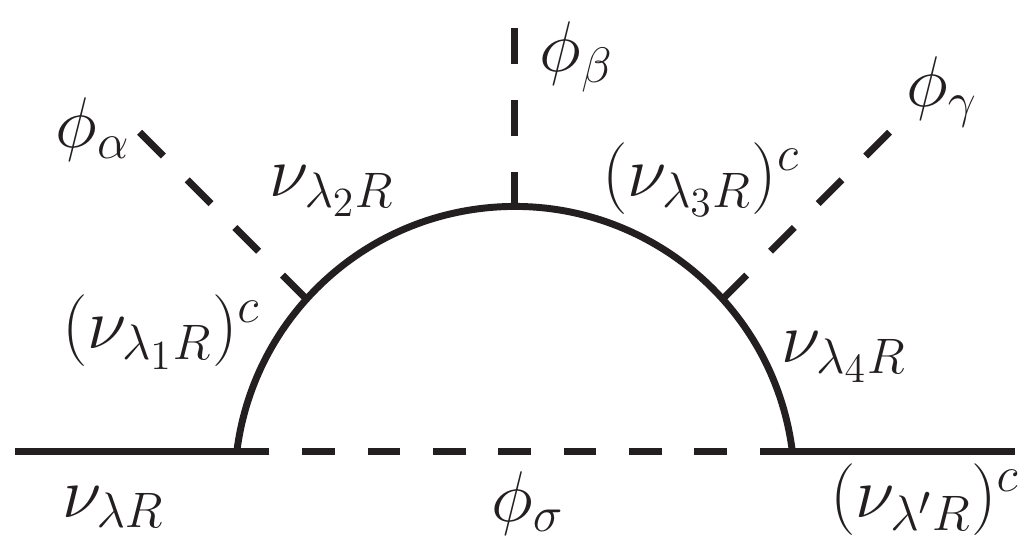} 
   \parbox{0.95\textwidth}{\vskip-6pt
   \caption{One-loop contributions to Majorana neutrino masses}
   \label{fig:3b}}
\end{minipage}
\end{figure}

For charged leptons, the Dirac masses are the entire masses, but it is not
necessarily true for neutrinos. ${\rm U(1)}_\lambda$ can be broken by
SM singlet $\phi_\lambda$, which cannot couple to the left-handed leptons, 
but can only couple to the right-handed neutrinos, 
whose couplings are necessarily Majorana types.
Then the Yukawa couplings are Majorana-type as
\beq
\label{e:s6}
\tilde{y}_{\lambda'\lambda}
\phi_\alpha \bar{\nu_{\lambda' R}}(\nu_{\lambda R})^c,
\eeq
and the Majorana masses for $\nu_R$'s are
\beq
\label{e:s6a}
\tilde{m}_{R\lambda'\lambda}\bar{\nu_{\lambda' R}}(\nu_{\lambda R})^c,
\eeq
which includes one-loop corrections according to
Fig.\ref{fig:3a} such that
\beq
\label{e:12}
\Delta {m}_{R\lambda'\lambda}^{(1)}
\sim
\sum_{\alpha,\beta,\gamma} {v_\alpha v_\beta v_\gamma\over \Lambda^2}\,
\tilde{y}_{\lambda\lambda'''}\, 
\tilde{y}_{\lambda'''\lambda''}\, \tilde{y}_{\lambda'\lambda''}\, 
\lambda_{\phi}.
\eeq
Fig.\ref{fig:3b} also contributes at the same order, but does not generate 
any new nontrivial contributions. In other words, if Fig.\ref{fig:3a} 
leads to vanishing components, so does Fig.\ref{fig:3b}.

\noindent
\underline{Physical Majorana Neutrino Masses for $\nu_i^M$}

Since the Dirac neutrino mass matrix $\mathbf{m}$ commutes with 
$\tilde{\mathbf{m}}_R$, it is easy to diagonalize to obtain physical
Majorana masses. The net neutrino mass terms are
\bseq
\begin{align}
\label{e:s14a}
&\ \bar{\nu_{\lambda' L}} m_{\lambda' \lambda}\nu_{\lambda R}
+\bar{(\nu_{\lambda' R})^c} m_{\lambda' \lambda}(\nu_{\lambda L})^c
+\bar{\nu_{\lambda' R}}\tilde{m}_{R\lambda' \lambda}(\nu_{\lambda R})^c \\
\label{e:s14c}
= &\ \bar{\nu_{i L}} m_{Di}\nu_{i R}
+\bar{(\nu_{i R})^c} m_{Di}(\nu_{i L})^c
+\bar{\nu_{i R}} \tilde{m}_{Ri}(\nu_{i R})^c \\
\label{e:s14d}
= &\ \bar{\nu_{i L}^M} m_{Li}(\nu_{i L}^M)^c
+\bar{\nu_{i R}^M} m_{Ri}(\nu_{i R}^M)^c,
\end{align}
\eseq
where
\beq
\label{e:s15a}
\begin{pmatrix}
\nu_{iR}^M \\
(\nu_{iL}^M)^c
\end{pmatrix}
=
\left(
\begin{array}{cc}
c_i & s_i \\ 
-s_i &c_i
\end{array}
\right)
\begin{pmatrix}
(\nu_{iL})^c \\
\nu_{iR}
\end{pmatrix}
\eeq
with
\bseq
\begin{align}
\label{e:s16a}
c_i &={m_{Di}\over\sqrt{m_{Ri}^2+m_{Di}^2}}
=-{m_{Li}\over\sqrt{m_{Li}^2+m_{Di}^2}}, \\
\label{e:s16b}
s_i &={m_{Ri}\over\sqrt{m_{Ri}^2+m_{Di}^2}}
={m_{Di}\over\sqrt{m_{Li}^2+m_{Di}^2}}, \\
\label{e:s16c}
m_{Li} &\equiv 
\half\left(\tilde{m}_{Ri}-\sqrt{\tilde{m}_{Ri}^2+4m_{Di}^2}\right),\\
\label{e:s16d}
m_{Ri} &\equiv 
\half\left(\tilde{m}_{Ri}+\sqrt{\tilde{m}_{Ri}^2+4m_{Di}^2}\right).
\end{align}
\eseq

\noindent
\underline{Yukawa coupling constants}

The values of Yukawa couplings depend on the masses and symmetry breaking 
scales. If physical neutrino masses are Dirac types, $y\ll 1$. If not, 
we can assume $\tilde{y}_{\lambda'\lambda}\sim \CO(1)$ for the maximum.
Since we expect that the left-handed Majorana masses are of order 
\beq
\label{e:15}
{m^2\over \tilde{m}_R}\sim 0.1\ {\rm eV},
\eeq
we can roughly estimate the Yukawa coupling constants $y_{\lambda\lambda}$ 
with respect to $v_\lambda$.
If we assume $v_\lambda\sim 1\ {\rm TeV}$, then
$m\sim 10^{11/2}\ {\rm eV}\sim 100\ {\rm MeV}$ 
so that we can obtain $y_{\lambda\lambda}\sim 10^{-3}$.
If we wish to get $y_{\lambda\lambda}\sim 1$ such that 
$m\sim 100\ {\rm GeV}$, then
$v_\lambda\sim \tilde{m}_R\sim 10^{11\times 2+1}\ {\rm eV} \sim 10^{14}\ 
{\rm GeV}$.
So, for $100\ {\rm MeV}\lsim m \lsim 10^{2}\ {\rm GeV}$,
$10^{-3}\lsim y_{\lambda\lambda}\lsim 1$. 

\noindent
\underline{Example I: $(q, 2q, -3q)$ with $\phi_q$ and $\phi_{2q}$}

First, consider $q_2=2q_1=2q$, and ${\rm U(1)}_\lambda$ is broken by two
singlet scalars, $\phi_q$ and $\phi_{2q}$,
then the ${\rm U(1)}_\lambda$ charge conservation on Yukawa couplings allows
$\mathbf{M}$ and $\mathbf{m}$ at one-loop order have only 
(23)-components vanishing,
while $\tilde{\mathbf{m}}_R$ has only the (22)-component vanishing.

For minimal mixing, i.e. $\tilde{s}\ll 1$, we can assume a low 
$v_\lambda \simeq 1\ {\rm TeV}$. Let $Y_{\ell\ell}\sim M_\ell/v_{\rm EW}$, 
$y\sim 10^{-3}$, $\tilde{y}\sim 1$, then we can obtain
$M_{12}\sim 10^{-4}\ {\rm MeV}$ and $M_{13}\sim 10^{-3}\ {\rm MeV}$ and that
$\tilde{s}_{13}\sim 10^{-6}\sim \tilde{s}_{12}$ and $\tilde{s}_{23}\sim 0$,
which leads to 
\bseq
\begin{align}
\label{e:16}
\mathbf{q} &\simeq
q\left(
\begin{array}{ccc}
1& -10^{-6}& 4\times 10^{-6}\\
 -10^{-6}&2 & 0\\
4\times 10^{-6} &0  &-3 
\end{array}
\right) \\
\label{e:16x}
V &\simeq\left(
\begin{array}{ccc}
1& -10^{-6}& -10^{-6}\\
 -10^{-6}&1 & 0\\
-10^{-6} &0  &1 
\end{array}
\right)
\end{align}
\eseq
such that
$
U_\nu= V U_R \simeq U_R.
$
With eq.(\ref{e:16}), the limit BR$(\mu\to 3e)<10^{-12}$ can constrain the 
${\rm U(1)}_\lambda$ breaking scale $v_\lambda$ as
$v_\lambda > v_{\rm EW}$.
So, our assumption $v_\lambda \sim 1\ {\rm TeV}$ is consistently valid 
with any choice of $q$.
The magnitude of $q$ cannot be directly constrained by any current 
data, but we can safely assume it is extra weak such that 
$q \ll g$, where $g$ is the Weak coupling constant.

As we increase $v_\lambda$ higher, we can generate larger off-diagonal masses
for charged leptons, which allows more significant cLFV. 
For example, for $v_\lambda\sim 10^{14}$ Gev, 
if $\tilde{s}_{13}\simeq -0.455$ and 
$\tilde{s}_{12}\simeq -0.973$ and $\tilde{s}_{23}\simeq 0.976$,
we can obtain 
\beq
\label{e:20}
\mathbf{q} \simeq
q\left(
\begin{array}{ccc}
0.923& 2&-0.24 \\
2 & -1.79& 0.73\\
-0.24  & 0.73 & 0.87
\end{array}
\right)
\eeq
with significant $\mu\to e$  cLFV.
However, we can even completely suppress $\mu\to e$ cLFV.
For $v_\lambda\sim 10^{14}$ Gev,
if $\tilde{s}_{13}\simeq -0.5$ and 
$\tilde{s}_{12}\simeq 0\simeq \tilde{s}_{23}$,
we can obtain 
\beq
\label{e:21}
\mathbf{q} \simeq
q\left(
\begin{array}{ccc}
0& 0&\sqrt{3} \\
0 & 2& 0\\
\sqrt{3}  & 0 & -2
\end{array}
\right).
\eeq
So we need additional conditions to make predictions more specific.

With some specific assumptions, we can make it more interesting. 
Suppose $y_{22}=y_{33}$ and
\beq
\label{e:22}
m_3 =s_{R12}^2 m_1+ c_{R12}^2 m_2,
\eeq
then
\bseq
\begin{align}
\label{e:23a}
\mathbf{M} &\simeq
\left(
\begin{array}{ccc}
83.42 & 43.73 &320.4\\ 
43.73 &83.42  &0\\
320.4 &0 &1716.15
\end{array}
\right)\ {\rm MeV},
\\
\label{e:23b}
\mathbf{q} &\simeq
q\left(
\begin{array}{ccc}
1.11& 0.467&0.638 \\
0.467 & 1.76& -0.357\\
0.638  & -0.357 & -2.86
\end{array}
\right).
\end{align}
\eseq
This leads to $s_{R13}=0$ for $U_R$. One notable fact in this case is that
the largest ratio in eq.(\ref{e:23a}) is about 40, but we can obtain
$M_\tau/M_e \sim 4\times 10^6$. 
In other words, we can obtain the large charged lepton
mass hierarchy without a large mass hierarchy in the mass matrix.
Furthermore, this is related to the smallness of $s_{13}$ in $U_\nu$.
In fact, we can do even better in the next example.

\noindent
\underline{Example II: $(q, 2q, -3q)$ with $\phi_{2q}$ and $\phi_{3q}$}

Consider the previous example, but replace $\phi_q$ with $\phi_{3q}$ and 
assume $y_{22}=0 \neq y_{33}$,
$M_{11} =M_{22}=M_\mu$, $\tilde{m}_{R23} =0$, and
\beq
\label{e:24}
\tilde{m}_{R3} = s_{R12}^2 \tilde{m}_{R1}+ c_{R12}^2 \tilde{m}_{R2}, 
\eeq 
then mass matrices are
\beq
\label{e:31}
\mathbf{M} =
\left(
\begin{array}{ccc}
M_\mu & 0 &M_{13}\\ 
0 &M_\mu  &0\\
M_{13} &0 &M_{33}
\end{array}
\right),\
\mathbf{m} =
\left(
\begin{array}{ccc}
m_{11} & m_{12} &m_{13}\\ 
m_{12} &0  &m_{23}\\
m_{13} &m_{23} &m_{33}
\end{array}
\right),\
\tilde{\mathbf{m}}_R =
\left(
\begin{array}{ccc}
\tilde{m}_{R11} & \tilde{m}_{R12} &\tilde{m}_{R13}\\ 
\tilde{m}_{R12} &\tilde{m}_{R22}  &0\\
\tilde{m}_{R13} &0 &\tilde{m}_{R22}
\end{array}
\right).
\eeq
Then eq.(\ref{e:10}) becomes
\beq
\label{e:25}
U_{\rm PMNS}=U_\nu(\theta_{23},\theta_{13},\theta_{12})
= V({\tilde{\theta}}_{13},{\tilde{\theta}}_{12}=0={\tilde{\theta}}_{23})\, 
U_R(\theta_{R23}, \theta_{R12}, \theta_{R13}=0),
\eeq
where
\bseq
\begin{align}
\label{e:26a}
V({\tilde{\theta}}_{13}) &= 
\left(
\begin{array}{ccc}
\tilde{c}_{13} & 0 &\tilde{s}_{13}\\ 
0 &1 &0\\
-\tilde{s}_{13} &0 &\tilde{c}_{13}
\end{array}
\right),\\
\label{e:26b}
U_R(\theta_{R12}, \theta_{R23}) &=
\left(
\begin{array}{ccc}
c_{R12} & s_{R12} &0\\ 
-c_{R23}s_{R12} &c_{R23}c_{R12} &s_{R23}\\
s_{R23}s_{R12} &-s_{R23}c_{R12} &c_{R23}
\end{array}
\right)
\end{align}
\eseq
and
\beq
\label{e:27}
\tilde{s}_{13}\equiv \sin\tilde{\theta}_{13}=\sqrt{M_\mu-M_e\over M_\tau -M_e}
\simeq \sqrt{M_\mu\over M_\tau}.
\eeq
With $s^2_{13}$ and $s^2_{12}$ given in eq.(\ref{e:s4bf}), eq.(\ref{e:25})
leads to $s_{23}^2\simeq 0.61$, $s^2_{R12}\simeq 0.50$, and 
$s^2_{R23}\simeq 0.59$.
So eq.(\ref{e:27}) is perfectly consistent with the best-fit neutrino 
mixing angles.
Due to $c_{23}=t_{13}/\tilde{t}_{13}$, 
as $s_{13}$ decreases or increases, other values
vary to the opposite direction. 
For an interesting example, let us choose $s_{13}^2\simeq 0.0296$, 
$s^2_{12}\simeq 0.328$, then eq.(\ref{e:25}) fixes $s^2_{23}\simeq 0.515$, 
$s^2_{R12}= 1/2= s^2_{R23}$, i.e. $U_R$ is bi-maximal\cite{Barger:1998ta}
such that $\tilde{m}_{R11}=\tilde{m}_{R22}$ and
$\tilde{m}_{R12}=\tilde{m}_{R13}$. 
This is within $2\sigma$, so still plausible.

Note that
\beq
\label{e:28}
s_{13}=\tilde{s}_{13} c_{R23} \simeq \sqrt{M_\mu\over M_\tau} c_{R23}
\eeq
justifies that the smallness of $s_{13}$ in terms of the smallness of 
the muon-to-tau mass ratio. The cLFV is based on
\beq
\label{e:29}
\mathbf{q} =
q\left(
\begin{array}{ccc}
\tilde{c}_{13}^2-3\tilde{s}_{13}^2& 0&-4\tilde{s}_{13}\tilde{c}_{13} \\
0 & 2& 0\\
-4\tilde{s}_{13}\tilde{c}_{13}  & 0 & \tilde{s}_{13}^2-3\tilde{c}_{13}^2
\end{array}
\right).
\eeq

\noindent
\underline{Example III: $(0, q, -q)$ with $\Phi_{q}$}

In fact, we can get eq.(\ref{e:25}) just for Dirac neutrinos,
hence without singlet $\phi_\lambda$.
Let $\mathbf{q}_d={\rm diag}(0,q,-q)$ and ${\rm U(1)}_\lambda$ be broken
by a hypercharge $1/2$ isospin doublet $\Phi_q$ at the same time as the EWSB.
This is allowed because $q$ can be sufficiently smaller than the Weak 
coupling in our model, and, above all, there is no experimental constraint 
against it.
With assumptions $Y_{12}=0$, $Y_{11}=Y_{22}$, $y_{22}=y_{33}$, 
the tree-level mass matrices are
\beq
\label{e:32}
\mathbf{M} =
\left(
\begin{array}{ccc}
M_\mu & 0 &M_{13}\\ 
0 &M_\mu  &0\\
M_{13} &0 &M_{33}
\end{array}
\right), \qquad
\mathbf{m} =
\left(
\begin{array}{ccc}
m_{11} & m_{12} &m_{13}\\ 
m_{12} &m_{22}  &0\\
m_{13} &0 &m_{22}
\end{array}
\right).
\eeq
Then we reproduce eqs.(\ref{e:25})-(\ref{e:27}) except
now subscript $R$ replaced by $D$,
In this case, the physical neutrinos are Dirac types and their masses satisfy
\beq
\label{e:33}
m_1+m_2=2m_3.
\eeq
The cLFV is based on
\beq
\label{e:34}
\mathbf{q} =
q\left(
\begin{array}{ccc}
-\tilde{s}_{13}^2& 0&-\tilde{s}_{13}\tilde{c}_{13} \\
0 & 1& 0\\
-\tilde{s}_{13}\tilde{c}_{13}  & 0 & -\tilde{c}_{13}^2
\end{array}
\right),
\eeq
where $e^+ e^-$ coupling to $Z^{(\lambda)}$ is relatively suppressed.

If $Y_{12}\neq 0$, but still $Y_{11}=Y_{22}$, we can also recover 
eqs.(\ref{e:23a})(\ref{e:23b}).

\noindent
\underline{Flavor Violating Higgs Decays}

\begin{figure}[t] 
\centering
\includegraphics[width=2in]{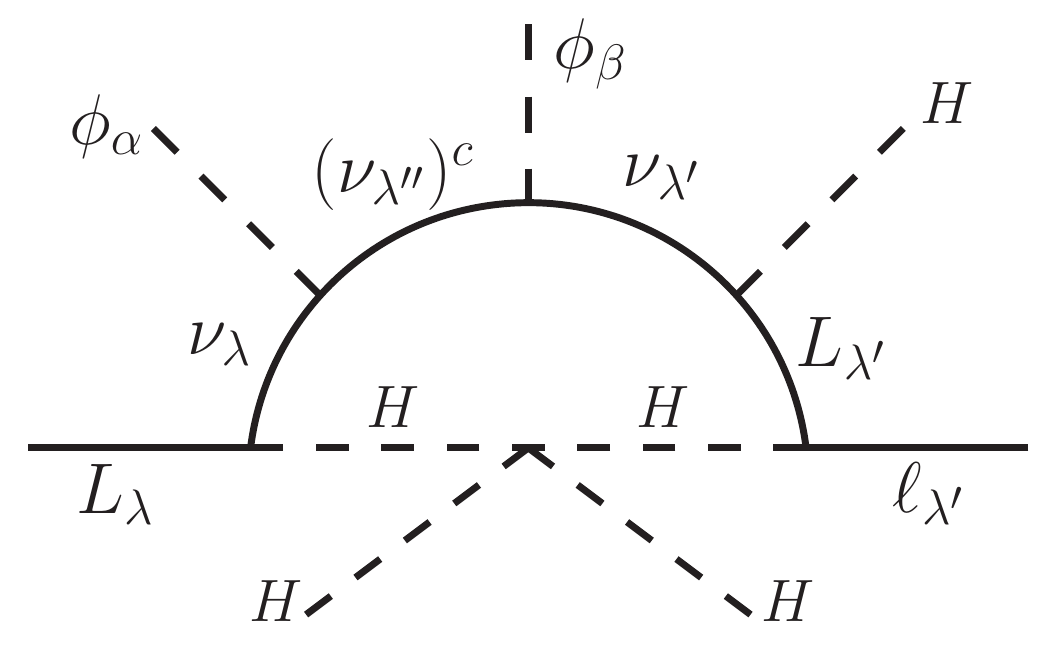} 
   \parbox{0.8\textwidth}{\vskip-6pt
   \caption{One-loop contributions to flavor-violating Yukawa couplings}
   \label{fig:1b}}
\end{figure}

Fig.\ref{fig:1b} shows further higher order contributions as
\beq
\label{e:r17}
\Delta M_{\lambda'\lambda}^{(2)}
\sim
\sum_{\alpha,\beta} {v_\alpha v_\beta v_{\rm EW}^3\over \Lambda^4}
Y_{\lambda'\lambda'}\, y_{\lambda\lambda}\, y_{\lambda'\lambda'}\, 
\tilde{y}_{\lambda\lambda''}\, \tilde{y}_{\lambda'\lambda''}\, 
\lambda_H.
\eeq
where $\lambda_H v_{\rm EW}^2 =m_H^2$.
This higher order may also generate flavor-violating Higgs decay
with Yukawa couplings\cite{Hall:1993ca}\cite{Harnik:2012pb}, 
provided it is not diagonal, of 
\beq
Y_{\ell'\ell} 
=V^*_{\lambda'\ell'}V_{\ell\lambda}
{3\Delta M_{\lambda'\lambda}^{(2)}\over v_{\rm EW}}.
\eeq

\noindent
\underline{Discussions}

The model we have proposed in this Letter is a truly minimal 
extension of the SM because it only needs the right-handed neutrinos
as additional matter accompanied by an extra abelian gauge boson.
The singlet scalar we need to break the extra ${\rm U(1)}_\lambda$ 
only interacts with the right-handed neutrinos in addition to the 
${\rm U(1)}_\lambda$ gauge boson, so well hidden from the current detections.
We can make the model more accessible with a SM doublet $\Phi_q$, 
which could be more easily testable.

This model makes interesting predictions on the LFV, which can be
distinguished from other models. 
For example, since the cLFV process of production
in our model is mainly caused by a gauge boson, compared with the cLFV 
based on Yukawa interactions, one should be able to distinguish our model 
by determining the spin of the boson in the charged lepton flavor 
violating processes whether the spin is one or zero.

We can construct models without $\sum q_\lambda = 0$ if the entire 
gauge symmetry is extended to that of the left-right symmetric 
models\cite{Mohapatra:1980yp}, as in \cite{La:2012ky}. 
In that case, ${\rm U(1)}_\lambda$ breaking should be above the 
spontaneous parity breaking scale.

It also has an attractive theoretical aspect beyond cLFV as a benefit of
having ${\rm U(1)}_\lambda$. 
The charged lepton mass hierarchy is more manageable since 
large $M_\tau/M_e$ mass ratio can be explained in terms of more 
comparable masses. This is possible because non-diagonal charged lepton
mass matrix can be constructed in the ${\rm U(1)}_\lambda$ basis,
which otherwise is hard to justify. This in turn links the charged lepton
masses to the neutrino mixing matrix. In particular,
the smallness of $\theta_{13}$ of the neutrino mixing can be related to
the smallness of the muon-to-tau mass ratio, and we can reproduce the 
neutrino mixing angles perfectly consistent with the latest data.
In this sense, our model is natural and 
we believe that this strongly suggests the worthiness of having 
${\rm U(1)}_\lambda$ family symmetry.

In fact, we can still obtain eqs.(\ref{e:25})-(\ref{e:27}) 
(or eq.(\ref{e:23a}) more generally)
just in the SM with $\nu_R$'s, if we allow charged leptons have different 
flavor and mass eigenstates. 
All we need is to replace $V$ with $U_\ell^\dagger$
and  $U_R$ with $U_\nu$, then demand eq.(\ref{e:31}) (eq.(\ref{e:23a})
respectively).
So it deserves further investigations 
in this direction, both theoretically and experimentally.

\noindent
{\bf Acknowledgments:}
I would like to thank Tom Weiler for many illuminating discussions and 
collaborating in part.

\renewcommand{\Large}{\large}


\begin{thebibliography}{100}
\setlength{\itemsep}{-1mm}

\bibitem{pdg}
K.~Nakamura and S.T.~Petcov, in Particle Data Group,  
  Phys.\ Rev.\ D {\bf 86}, 010001 (2012)
(http://pdg.lbl.gov/2012/reviews/rpp2012-rev-neutrino-mixing.pdf)
and references therein.

\bibitem{Pontecorvo:1957qd} 
  B.~Pontecorvo,
  Sov.\ Phys.\ JETP {\bf 7}, 172 (1958)
  [Zh.\ Eksp.\ Teor.\ Fiz.\  {\bf 34}, 247 (1957)].

\bibitem{no_ref}
B. Pontecorvo, Sov. Phys. JETP 6, (1957) 429; 
Zh. Eksp. Teor. Fiz. 33 (1957) 549;
Z. Maki, M. Nakagawa and S. Sakata, Prog. Theor. Phys. 28 (1962 ) 870; 
B. Pontecorvo, Sov. Phys. JETP 26 (1968) 984; 
Zh. Eksp. Teor. Fiz. 53 (1968) 1717; 
V. N.Gribov and B. Pontecorvo, Phys. Lett. B 28 (1969) 493.

\bibitem{Hoecker:2012np} 
  A.~Hoecker,
  Pramana {\bf 79}, 1141 (2012)
  [arXiv:1201.5093 [hep-ph]].

\bibitem{Deppisch:2012vj} 
  F.~F.~Deppisch,
  arXiv:1206.5212 [hep-ph].

\bibitem{La:2012ky} 
  H.~La,
  arXiv:1209.1377 [hep-ph].

\bibitem{Froggatt:1978nt} 
  C.~D.~Froggatt and H.~B.~Nielsen,
  Nucl.\ Phys.\ B {\bf 147}, 277 (1979).

\bibitem{Foot:1989ts} 
  R.~Foot, G.~C.~Joshi and H.~Lew,
  Phys.\ Rev.\ D {\bf 40}, 2487 (1989).

\bibitem{FileviezPerez:2010gw} 
  P.~Fileviez Perez and M.~B.~Wise,
  Phys.\ Rev.\ D {\bf 82}, 011901 (2010)
  [Erratum-ibid.\ D {\bf 82}, 079901 (2010)]
  [arXiv:1002.1754 [hep-ph]].

\bibitem{Chao:2010mp} 
  W.~Chao,
  Phys.\ Lett.\ B {\bf 695}, 157 (2011)
  [arXiv:1005.1024 [hep-ph]].

\bibitem{Schwaller:2013hqa} 
  P.~Schwaller, T.~M.~P.~Tait and R.~Vega-Morales,
  arXiv:1305.1108 [hep-ph].

\bibitem{Langacker:2008yv} 
  P.~Langacker,
  Rev.\ Mod.\ Phys.\  {\bf 81}, 1199 (2009)
  [arXiv:0801.1345 [hep-ph]].

\bibitem{Langacker:2009im} 
  P.~Langacker,
  AIP Conf.\ Proc.\  {\bf 1200}, 55 (2010)
  [arXiv:0909.3260 [hep-ph]].

\bibitem{Holdom:1985ag} 
  B.~Holdom,
  Phys.\ Lett.\ B {\bf 166}, 196 (1986).

\bibitem{Ibanez:1994ig} 
  L.~E.~Ibanez and G.~G.~Ross,
  Phys.\ Lett.\ B {\bf 332}, 100 (1994)
  [hep-ph/9403338].

\bibitem{Demir:2005ti} 
  D.~A.~Demir, G.~L.~Kane and T.~T.~Wang,
  Phys.\ Rev.\ D {\bf 72}, 015012 (2005)
  [hep-ph/0503290].

\bibitem{Langacker:2000ju} 
  P.~Langacker and M.~Plumacher,
  Phys.\ Rev.\ D {\bf 62}, 013006 (2000)
  [hep-ph/0001204].

\bibitem{Chen:2006hn} 
  M.~-C.~Chen, A.~de Gouvea and B.~A.~Dobrescu,
  Phys.\ Rev.\ D {\bf 75}, 055009 (2007)
  [hep-ph/0612017].

\bibitem{pdg2013}
Particle Data Group,
http://pdg8.lbl.gov/rpp2013v2/pdgLive/

\bibitem{Barger:1998ta} 
  V.~D.~Barger, S.~Pakvasa, T.~J.~Weiler and K.~Whisnant,
  Phys.\ Lett.\ B {\bf 437}, 107 (1998)
  [hep-ph/9806387].

\bibitem{Hall:1993ca} 
  L.~J.~Hall and S.~Weinberg,
  Phys.\ Rev.\ D {\bf 48}, 979 (1993)
  [hep-ph/9303241].

\bibitem{Harnik:2012pb} 
  R.~Harnik, J.~Kopp and J.~Zupan,
  JHEP {\bf 1303}, 026 (2013)
  [arXiv:1209.1397 [hep-ph]].

\bibitem{Mohapatra:1980yp} 
  R.~N.~Mohapatra and G.~Senjanovic,
  Phys.\ Rev.\ D {\bf 23}, 165 (1981).

\end{thebibliography}
\end{document}